\documentclass[12pt]{article}
\renewcommand{\baselinestretch}{1.5}
\usepackage{amssymb,bm,amsmath,amsthm}
\usepackage{pdfpages}
\usepackage{listings}
\usepackage{graphicx}
\usepackage{enumerate}
\usepackage{authblk}
\usepackage{tikz}

\newtheorem{theorem}{Theorem}

\theoremstyle{remark}
\newtheorem{remark}{Remark}

\theoremstyle{definition}
\newtheorem{application}{Application}

\newcommand{\nc}{\newcommand}
\newcommand\Wtilde[1]{\stackrel{\sim}
{\smash{#1}\rule{0pt}{1.1ex}}}
\nc{\R}{\mathbb R}
\nc{\cov}{\mathcal{C} ov}
\graphicspath{{figure/}}

\usepackage{natbib}
\usepackage[toc,page]{appendix}
\usepackage{multicol, thmtools, xcolor}
\usepackage{graphicx, mathtools}
\definecolor{watpink}{RGB}{198, 0, 120}
\definecolor{watyellow}{RGB}{254, 211, 76}
\definecolor{watyellow2}{RGB}{235,171, 0}
\definecolor{skyblue}{RGB}{0, 102, 204}

\nc{\red}[1]{\textcolor{red}{#1}}
\nc{\blue}[1]{\textcolor{blue}{#1}}
\nc{\green}[1]{\textcolor{green}{#1}}
\nc{\purple}[1]{\textcolor{purple}{#1}}
\nc{\violet}[1]{\textcolor{violet}{#1}}
\nc{\wpink}[1]{\textcolor{watpink}{#1}}
\nc{\btheta}{\boldsymbol \theta}
\nc{\bI}{\boldsymbol{\mathrm{I}}}
\nc{\bx}{\boldsymbol{\mathrm{x}}}
\nc{\bff}{\boldsymbol{\mathrm{f}}}
\DeclareMathOperator{\tr}{tr}
\DeclareMathOperator{\Eff}{\mathbb{E}ff}

\DeclarePairedDelimiter\abs{\lvert}{\rvert}%
\usepackage[colorlinks = true, 
linkcolor = watpink, 
citecolor = skyblue, 
final]{hyperref}
\usepackage{orcidlink}
\nc{\yck}[1]{\textcolor{red}{#1}}

\allowdisplaybreaks
\tikzset{every picture/.style={line width=0.75pt}} 

\begin{document}

\def\spacingset#1{\renewcommand{\baselinestretch}%
	{#1}\small\normalsize} \spacingset{1}

{\small
	\title{\bf CVXSADes: a stochastic algorithm for constructing optimal exact regression designs with single or multiple objectives}
	\author[1]{Chi-Kuang Yeh \orcidlink{0000-0001-7057-2096}\thanks{Corresponding author: chi-kuang.yeh@uwaterloo.ca}}
	\author[2]{Julie Zhou}
	\affil[1]{Department of Statistics and Actuarial Science, University of Waterloo}
	\affil[2]{Department of Mathematics and Statistics, University of Victoria}

    \date{}
	\setcounter{Maxaffil}{0}
	\renewcommand\Affilfont{\itshape\small}
	\maketitle
}

\begin{center}
{\small
\today}
\end{center}

\bigskip

\noindent
\begin{center}
    {\footnotesize ABSTRACT}    
\end{center}

{\footnotesize  
We propose an algorithm to construct optimal \textit{exact designs} (EDs). Most of the work in the optimal regression design literature focuses on the \textit{approximate design} (AD) paradigm due to its desired properties, including the optimality verification conditions derived by \citet{kiefer:1959:optimum,kiefer:1974:equivalence-theory}. ADs may have unbalanced weights, and practitioners may have difficulty implementing them with a designated run size $n$.  Some EDs are constructed using rounding methods to get an integer number of runs at each support point of an AD, but this approach may not yield optimal results. To construct EDs, one may need to perform new combinatorial constructions for each $n$, and there is no unified approach to construct them. Therefore, we develop a systematic way to construct EDs for any given $n$. Our method can transform ADs into EDs while retaining high statistical efficiency in two steps. The first step involves constructing an AD by utilizing the convex nature of many design criteria. The second step employs a simulated annealing algorithm to search for the ED stochastically.  Through several applications, we demonstrate the utility of our method for various design problems. Additionally, we show that the design efficiency approaches unity as the number of design points increases.
\vspace{0.5cm}


}
\newpage

\section{Introduction}\label{sec-introduction}

Consider a general regression model,
\begin{equation}\label{M1}
y_i=\eta(\bx_i, \btheta)+ \epsilon_i, \quad i=1, \ldots, n,
\end{equation}
where $y_i$ is the $i$-th observation of a response variable $y$ at design point
$\bx_i \in S \subset \R^p$, $S$ is a design space,
$\btheta \in \R^q$ is the unknown regression parameter vector,
response function $\eta(\bx_i, \btheta)$   can be a linear or nonlinear
function of $\btheta$, and the errors $\epsilon_i$ are assumed to be uncorrelated with mean zero and finite variance $\sigma^2$.
Let $\hat{\btheta}$ be an estimator of $\btheta$, such as the least squares estimator. Various optimal designs are defined by minimizing
$\phi\left\{ \cov(\hat{\btheta}) \right\}$ over the design points
$\bx_1, \ldots, \bx_n$, where function $\phi(\cdot)$  
can be determinant, trace, or other scalar functions. The resulting designs are called optimal exact
designs (OEDs), which depend on the response function $\eta(\cdot,\cdot)$, the design space $S$, 
the estimator $\hat{\btheta}$, the scalar function $\phi(\cdot)$, and the number of points $n$.
As for searching for the OEDs, coordinate exchange and simulated annealing (SA) algorithms 
have been developed and used; see 
\citet{meyer:1988, meyer:1995:coordinate-exchange}, \citet{wilmut:2011:two-level-ffd}, \citet{smucker:2012:coordinate-exchange}, \citet{rempel:2014:K-opt}
and \citet{Cuervo:2016:exchange}, for a small sample of recent contributions to this problem. It is well known that it is difficult to construct OEDs, even for relatively simple problems; see Section 1.7 in \citet{berger-wong:2009} for more details. Note that other than the model and the parameter values, the exact designs also depend on the number of the run size $n$, and the practitioner needs to recalculate the design for each $n$, which makes it challenging to construct in practice.

To avoid calculating a huge number (near-infinite) of  different exact designs for each given $n$, \citet{kiefer:1959:optimum,kiefer:1974:equivalence-theory} proposed and developed the general equivalence theory for \textit{optimal approximate designs} (OADs). With the approximate design, one does not need to recalculate the design for each $n$. Instead, it calculates the proportion of how many resources should be allocated at different support points. The equivalence theorem is useful for constructing OADs analytically and numerically. After obtaining OADs, we may use rounding to convert them to exact designs. This method is usually suggested in research papers; e.g. \citet{pukelsheim:92:rounding} and its follow-up works. Before explaining the details of the conversion, we first give a short review of OADs. Let $\xi(\bx)$  be a discrete distribution (design) with $k$ support points in $S$, say, ${\bf v}_1, \ldots, {\bf v}_k$, and their corresponding weights are denoted by, $w_1, \ldots, w_k$, respectively. 
Note that $k$ is not fixed and can be any positive integer.
Denote the set of all discrete distributions on $S$ as $\Xi_S$.
The information matrix of a design $\xi(\bx)\in \Xi_S$ for model \eqref{M1} is given by
\begin{eqnarray}
\bI(\xi, \btheta^*) =\sum_{i=1}^k w_i \bff({\bf v}_i, \btheta^*)
\bff^\top({\bf v}_i, \btheta^*),
\label{eq-info1}
\end{eqnarray}
where vector $\bff(\bx, \btheta) =\frac{\partial \eta(\bx, \btheta)}
{\partial  \btheta}$, and $\btheta^*$ is the true value of $\btheta$.
The covariance matrix of $\hat\btheta$,
$\cov(\hat\btheta)$,
 is proportional to $\bI^{-1}(\xi, \btheta^*)$.
An OAD is defined as the minimizer of  $ \phi\left\{  \bI^{-1}(\xi, \btheta^*)   \right\} $ over all possible designs  $\xi \in \Xi_S$ for a given function $\phi(\cdot)$. If $\bI(\xi, \btheta^*) $ depends on $\btheta^*$, then the OAD is called  a \textit{locally OAD} or simply    
OAD in this paper.   
Notice that, for linear response functions, $\bI(\xi, \btheta^*) $ does not depend on
$\btheta^*$. In practice, we do not know $\btheta^*$, and replace $\btheta^*$ in \eqref{eq-info1} by an estimate, which may be available from pilot studies or the domain knowledge.

Since it is easier to construct OADs than OEDs, 
OADs have been obtained for various models, design spaces, and optimality
criteria.
Often numerical methods are used for finding OADs, and the methods
include, for example,  multiplicative algorithm \citep{zhang-mukerjee:2013,bose-mukerjee:2015:assymetric},  
cocktail algorithm \citep{yu:2011:cocktail-algorithm},
genetic algorithm \citep{broudiscou:1996:design-genetic,hamada:2001:genetic-bayesian}, semi-definite programming method \citep{papp:2012:design-rational,duarte-wong:2015:SDP,ye:2017:SDP}, semi-infinite programming tools in \citet{duarte:2014:SIP} and \citet{duarte:2015:SIP}, particle swarm method \citep{chen:2015:minimax}, convex optimization method via CVX toolbox \citep{grant:2020:cvx-guide} in \citet{gao:2017:design-second-order}, a general method
in \citet{yang:2013:design-algorithm}, and an efficient method in \citet{duan:2022:design-computation}. \citet{mandal:2015:algorthimic} provides a comprehensive review of the algorithmic approaches utilized in most of the methods mentioned above. Recently,
\citet{wong:2019:cvx,wong:2023:CVX} gave detailed discussions and comments on several algorithms,
and  they also developed effective numerical algorithms for finding OADs
with multiple objective functions. 

As the advantages of OADs are mentioned, tremendous efforts have been put into them, but how to effectively construct $n$ design points efficiently from those OADs is still uncertain. Suppose $\xi_{\phi}^*$ is an OAD with $m$ support points in $S$, say ${\bf v}_1^*, 
\ldots, {\bf v}_m^*$, and their weights are $w_1^*, \ldots, w_m^*$, respectively. 
It is clear that  $\xi_{\phi}^*$ does not depend on $n$.  However, it may not be easy to implement
$\xi_{\phi}^*$ with $n$ runs in practice, as $nw_i^*$, $i=1, \ldots, m$, are usually not integers.
A general
suggestion is to round each $nw_i^*$ to the nearest positive integer subject to 
the total number of design points being $n$ \citep{wong:2019:cvx}.   
In some situations, how best to round $nw_i^*$ may not be clear.
For instance,  if an
approximate design has weights as
$(w_1^*, \ldots, w_5^*)=(0.2493,0.2465,0.1033,0.1517,0.2492)$, 
from a design in \citet[Example 4.2]{haines:2018:two-logistic},
then it is not clear how to choose $n=10$ design points from the approximate design. Are there good strategies for selecting $n$ design points from OADs?

In addition, the debate surrounding this rounding method has persisted for a while.  In a review paper by \citet{john:1975:rounding-error}, the author stated, ``the rounding-off procedure may eliminate points with small measure, thereby changing the nature of the design." \citet{lopez:2023:book} recalls in his recent book, stating that the controversy regarding exact design is discussed in Section 1.4.4, where he mentions that the rounding approach from approximate design to exact design is highly controversial, with some, including Box, dissenting. Particularly, on page 7, \citet{lopez:2023:book} states, ``Box has never accepted the use of the approximate designs introduced by \citet{kiefer:1959:optimum}," and on page 11, he further remarks,
``This idea came from \citet{kiefer:1974:equivalence-theory} and it used to be a controversial topic that George Box and others never have liked." In particular, when the number of available runs subject to the available resources is small, it is unclear which support points to keep from the approximate design or whether to keep any of them at all. \citet{mukerjee:2016:baseline-parameterization} also raised issues related to rounding for fractional factorial designs. They studied procedures for finding highly efficient exact designs from approximate design, however their procedure can only be applied for designs with a finite number of points.

In this paper, we propose a stochastic algorithm to construct OEDs to address these issues and systematically construct exact designs. Our proposed method utilizes a meta-heuristic algorithm, which does not rely on restrictive assumptions and does not require computing the gradient of the loss (objective) function in optimal design problems, which may be difficult to obtain in complicated design problems. Our algorithm first constructs an OAD and uses it as a starting point to search for exact designs. Additionally, OADs are used to compute the design efficiency of OEDs. Various optimality criteria are studied, including single-objective and multiple-objective criteria. To summarize, this paper makes three important contributions to optimal design of experiments as follows.

\begin{description}
    \item{(i)} \textbf{Fast and gradient-free algorithm}: Our first contribution is the development of a general algorithm for finding highly efficient OEDs via a CVX solver and an SA algorithm. Highly efficient OEDs can be found easily from the algorithm. In particular, our method does not require the derivative/gradient and Hessian matrix of the objective function of the design problems. 
\item{(ii)} \textbf{Importance of finding both exact  and approximate designs}: Our second contribution is to use our algorithm for finding both the OAD and OED. We show the importance of searching for both OAD and OED together. If we just use a SA algorithm to search for exact
designs, we do not know if the resulting designs are optimal or efficient. Computing both OED and OAD allows us to compute the design efficiency of an OED.
 If we just search for OADs and use a rounding method to obtain OEDs, then the resulting designs may not be efficient, or we do not know how to do the rounding. We can find highly efficient exact designs by computing an OAD and then searching for an exact design. The efficiency is computed by using the
OAD.
\item{(iii)} \textbf{Exact design on complex setup}: Our third contribution is to construct highly efficient exact designs in complicated settings. The conventional methods to compute the exact designs in the literature mainly centre around the low dimensional (small $p$) problems and one objective function; for instance, \citet{duarte:2020:exact-MINP} only considers the settings in at most three dimensions and only one objective function. There is a need to fill the gap between constructing the exact design in more complex settings, including design in high-dimension, and for more than one objective function. Our method is demonstrated to be able to find exact designs in higher dimensions, seven-dimensional space, and exact designs with multiple competing objectives. We present four applications with various optimality criteria and design spaces, including multiple-objective criteria and high-dimensional design spaces.
\end{description}

The rest of the paper is organized as follows. In Section \ref{sec-criteria}, we review the optimal design criteria with single and multiple objectives, related equivalence theory for OADs, and convex optimal design problems on discrete design spaces. In Section \ref{sec-algorithm}, we develop an effective algorithm to find OADs and OEDs for single and multiple objectives. We also give several properties about the OADs and OEDs and many remarks on the properties of the proposed algorithm.
Section \ref{sec-application} presents several applications and their OEDs, which are difficult to find by conventional methods.
Finally, we close the paper with the concluding remarks in Section \ref{sec-conclusion}. All proofs and derivations are in the Appendix. The implementation is available on the author's GitHub page \texttt{\href{https://github.com/chikuang/CVXSADes}{https://github.com/chikuang/CVXSADes}}.

\section{Optimality criteria with single or multiple objectives }\label{sec-criteria}

Various optimality criteria have been proposed and studied in the literature;
see, for example, \citet{fedorov:1972}, \citet{pukelsheim:1993}, \citet{berger-wong:2009}, and \citet{dean:2015:design-handbook}.  Here 
we recall several criteria to illustrate optimal design problems and equivalence theory,
and present convex optimization problems on discrete design spaces.

\subsection{Optimal design problems}

A-, c-, D-, I-, and E-optimality criteria are commonly used to construct optimal designs with a single
objective function.  The optimal design problems for finding
OADs can be written as
\begin{eqnarray}
\min_{\xi \in \Xi_S} \phi\left\{ \bI^{-1}(\xi, \btheta^*)   \right\},
\label{ProbOne}
\end{eqnarray}
where $\bI(\xi, \btheta^*)$ is defined in \eqref{eq-info1},
and $\phi$ is a scalar function.  For example,  
$\phi$ is the determinant function for D-optimality, and trace for A-optimality. We denote trace and determinant functions as $\tr(\cdot)$ and $\det(\cdot)$, respectively.
Let $\xi_{\phi}^*$ be the solution to problem \eqref{ProbOne}, which depends on $\phi$,
and $\xi_{\phi}^*$ is called an OAD.

The equivalence theory states the necessary and sufficient condition that an OAD satisfies. A general form for the condition is
 \begin{eqnarray}
d_{\phi}(\bx,\xi_{\phi}^*) \le 0, ~~\mbox{for all}~\bx \in S,
\label{Cond0}
\end{eqnarray}
where function $d_{\phi}(\bx,\xi_{\phi}^*) $ depends on the optimality criterion and is the negative of the directional derivative of $\phi$.
The equality in \eqref{Cond0} holds at the support points of $\xi_{\phi}^*$.
For a D-optimal design $\xi_{\phi}^*$,  
\begin{eqnarray*}
d_{\phi}(\bx,\xi_{\phi}^*)=\bff^\top(\bx, \btheta^*)
\bI^{-1}(\xi_{\phi}^*, \btheta^*) 
\bff(\bx, \btheta^*) - q.
\label{CondD}
\end{eqnarray*}
If $\phi\left\{ \bI^{-1}(\xi, \btheta^*)   \right\}=\tr 
\left\{{\bf C}^\top \bI^{-1}(\xi, \btheta^*)  {\bf C} \right\} $ with a constant matrix
${\bf C}$ ($q \times r$; $r \le q$),
which includes A-, c-, and I-optimality criteria,
then 
\begin{eqnarray*}
d_{\phi}(\bx,\xi_{\phi}^*)=\bff^\top(\bx, \btheta^*)
\bI^{-1}(\xi_{\phi}^*, \btheta^*) {\bf C}{\bf C}^\top
\bI^{-1}(\xi_{\phi}^*, \btheta^*) 
\bff(\bx, \btheta^*) -  \tr 
\left\{{\bf C}^\top \bI^{-1}(\xi_{\phi}^*, \btheta^*)  {\bf C} \right\}.
\label{CondAC}
\end{eqnarray*}

For a generalized linear model (GLM), the maximum likelihood estimator is often used to estimate parameter $\btheta$.
The information matrix of design $\xi(\bx)$ is different from 
that in \eqref{eq-info1} and  can be written as
\begin{equation}\label{eq-info2}
\bI(\xi, \btheta^*) =\sum_{i=1}^k w_i 
\lambda({\bf v}_i, \btheta^*) \bff({\bf v}_i)
\bff^\top({\bf v}_i),
\end{equation}
where $\lambda(\bx, \btheta^*)$ and $\bff(\bx)$ depend on the link  
and predictor functions of the GLM. 
In function $d_{\phi}(\bx,\xi_{\phi}^*)$,
we replace $\bff(\bx, \btheta^*)$ by
$\sqrt{\lambda({\bx}, \btheta^*)} \bff(\bx)$, and several examples are given in Section \ref{sec-application}.

\subsection{Optimal design problems on discrete design spaces}

The necessary and sufficient condition in \eqref{Cond0} enables us to find 
analytical and numerical solutions of $\xi_{\phi}^*$ for various models, including polynomial, second-order, and nonlinear models.  However, it is still challenging to find
$\xi_{\phi}^*$ for complicated models or  design spaces.
\citet{wong:2019:cvx} discussed and investigated OADs on discrete 
design spaces.  They used the fact that optimal design problems are convex optimization 
problems for commonly used optimality criteria, and applied CVX solver for finding 
OADs. Here we recall some details of the optimal design problems
on discrete design spaces.

Let $S_N =\{{\bf u}_1, \ldots, {\bf u}_N\} \subset S$ be a discrete design space with $N$ points,
where ${\bf u}_1, \ldots, {\bf u}_N$ are user selected points. One possible choice is to
use equally spaced grid points in $S$. Denote any distribution on $S_N$ by
$$\xi_N =\left( \begin{array}{cccc}
 {\bf u}_1 & {\bf u}_2 & \ldots &{\bf u}_N  \\
 w_1 & w_2 & \ldots & w_N  \\
 \end{array}
 \right), $$ 
 where the weights satisfy $w_j \ge 0$ for $j=1, \ldots, N$ and $\sum_{j=1}^N w_j=1$, and
 a point ${\bf u}_j$ is a support point of $\xi_N$ if the corresponding weight $w_j > 0$. 
 Let $\Xi_{S_N}$ be the set of all distributions on $S_N$.
The information matrix of design $\xi_N$ become
$\bI(\xi_N, \btheta^*)$, which can be computed from \eqref{eq-info1} or
\eqref{eq-info2},  replacing ${\bf v}_i$ by ${\bf u}_i$ and replacing $\sum_{i=1}^k$ by
$\sum_{i=1}^N$.
It is important to notice that in $\xi_N$, ${\bf u}_1, \ldots, {\bf u}_N$  are fixed points, but
$w_1, \ldots, w_N$ are unknown weights.  In addition,
$\bI(\xi_N, \btheta^*)$ is linear in weights $w_1, \ldots, w_N$.  Let weight vector
${\bf w}=(w_1, \ldots, w_N)^\top$.  Then
$\phi\left\{ \bI^{-1}(\xi_N, \btheta^*)   \right\}$
is a convex function of ${\bf w}$ for commonly used optimality criteria; see
\citet{boyd:2004:convex}, and \citet{wong:2019:cvx}.
An OAD on $S_N$, denoted by
$\xi_{\phi,N}^*$, is a solution to the following optimization problem,
$$\min_{\xi_N \in \Xi_{S_N}} \phi\left\{ \bI^{-1}(\xi_N, \btheta^*)   \right\},$$
or
\begin{eqnarray}
\left\{  \begin{array}{l}
\min_{\bf w} \phi\left\{ \bI^{-1}(\xi_N, \btheta^*)   \right\} \\
\mbox{subject to:} ~~w_j \ge 0,  ~j=1, \ldots, N,~~\sum_{j=1}^N w_j=1.
\\
\end{array} \right.
\label{eq-OptSN}
\end{eqnarray}
Problem \eqref{eq-OptSN} is a constraint convex optimization problem if $
\phi\left\{ \bI^{-1}(\xi_N, \btheta^*)   \right\}$ is a convex function of ${\bf w}$.
It can be solved by CVX solver, and detailed procedures of using CVX for finding 
$\xi_{\phi,N}^*$ are given in 
\citet{wong:2019:cvx}. 


\subsection{Multiple-objective optimal designs }\label{ssec-multiple-obj}

For multiple-objective optimal designs, there are mainly three optimality criteria, which are often
used.  They are compound, multiple efficiency constraint, and maximin efficiency criteria \citep{wong:2023:CVX}.  Since compound and multiple efficiency constraint criteria need extra information
to form the design problems, we only focus on maximin efficiency criterion to discuss OEDs.
Let $\phi_1(\xi), \ldots, \phi_\ell(\xi)$ be $\ell$ objective functions.  For instance,
$\phi_1(\xi)=\det\left\{ \bI^{-1}(\xi, \btheta^*) \right\}, \ldots,
\phi_\ell(\xi)=\tr \left\{\bI^{-1}(\xi, \btheta^*)\right\}$,
which are defined as different scalar functions of the same information matrix of a model.
They can also be defined as the same scalar function of information matrices of several models.
Suppose there are three competing models for an experiment, which leads to three different
information matrices, say, $\bI_1(\xi, \btheta_1^*)$, 
$\bI_2(\xi, \btheta_2^*)$,
and $\bI_3(\xi, \btheta_3^*)$, where $\btheta_1^*, \btheta_2^*$,
and $\btheta_3^*$ are the true parameters for the three models, respectively.
We may then define
$\phi_1(\xi)=\det\left\{ \bI_1(\xi, \btheta_1^*)\right\}$,
$\phi_2(\xi)=\det\left\{ \bI_2(\xi, \btheta_2^*)\right\}$, and
$\phi_3(\xi)=\det\left\{ \bI_3(\xi, \btheta_3^*)\right\}$.
Alternatively, we can use different scalar functions of those information matrices.

A maximin efficiency design is defined below. First, we minimize each $\phi_i(\xi)$ over $\xi$ to obtain an OAD $\xi_{\phi_i}^*$, $i=1, \ldots, \ell$.  Second, we
define design efficiency for a design $\xi$ and a given criterion $\phi$ as
\begin{eqnarray}
\Eff_{\phi}(\xi) =   \frac{\phi\left\{ \bI^{-1}(\xi_{\phi}^*, \btheta^*) \right\}}
{\phi\left\{ \bI^{-1}(\xi, \btheta^*) \right\}}.
\label{EffA}
\end{eqnarray}
For the $\ell$ objective functions, there are $\ell$ efficiencies  for a design $\xi$, which can
be written as
$$\Eff_{\phi_i}(\xi) =\frac{\phi_i(\xi_{\phi_i}^*)}{\phi_i(\xi)}, ~~i=1, \ldots, \ell.$$
Third, we find a solution to the maximin problem given as
\[
    \max_{\xi \in \Xi_S} \min_{1 \le i \le l} ~\Eff_{\phi_i}(\xi),
\]
and the solution is called a maximin efficiency design. The maximin problem is hard to solve. However, this problem can be transformed into a convex optimization problem on a discrete design space $S_N$, which can be solved easily via CVX solver.  
On $S_N$, in the maximin problem we replace $\xi$ and $\Xi_{S}$ by $\xi_N$ and $\Xi_{S_N}$,
respectively, 
and use $\xi_{\phi_i,N}^*$ to compute the efficiencies. \citet{wong:2023:CVX} developed an effective algorithm for finding maximin optimal designs with
various kinds of objective functions. \citet{gao:2024:multiple-objective} also derived the necessary and sufficient
conditions for multiple-objective optimal designs on $S_N$.

\section{Algorithms for OEDs}\label{sec-algorithm}
In this section, we develop an algorithm to construct OEDs which depends on $n$. Let $\xi_{n,\phi}^*$ be the OED for  an optimality criterion $\phi$, and its support points  are denoted by
${\tilde \bx}_1, \ldots, {\tilde \bx}_m$ with corresponding weights
${\tilde w}_1, \ldots, {\tilde w}_m$. In contrast to $u_1,\cdots,u_N$, these ${\tilde \bx}_1$ are not fixed/user specified. The weights must satisfy (i) 
$n{\tilde w}_i $ is a positive integer  for each $i=1, \ldots, m$, (ii) $\sum_{i=1}^m {\tilde w}_i=1$. Define $\Xi_n$ to be the set of all exact designs on $S$ with size $n$. It is clear that $\Xi_n \subset \Xi_s$. 
Since $\xi_{n,\phi}^* \in \Xi_n \subset \Xi_S$ and the OAD $\xi_{\phi}^*$ is a solution
to \eqref{ProbOne}, it is obvious that
$$ \phi\left\{ \bI^{-1}(\xi_{n,\phi}^*, \btheta^*) \right\}  
\ge \phi\left\{ \bI^{-1}(\xi_{\phi}^*, \btheta^*) \right\},$$
which implies that $\Eff_{\phi}(\xi_{n,\phi}^*) \le 1$.
A design $\xi$ is said to be highly efficient if $\Eff_{\phi}(\xi)$ is close to $1$.
In practice we may use 
$\Eff_{\phi}(\xi) \ge 0.95$ to define highly efficient
design $\xi$.

For D-optimality, we may minimize
$\det\left\{ \bI^{-1}(\xi, \btheta^*) \right\}$
or $\log \left(\det\left\{ \bI^{-1}(\xi, \btheta^*) \right\} \right)$
to find D-optimal designs, but the 
D-efficiency is given by
$$\Eff_{D}(\xi) =   \frac{\left(
\det \left\{ \bI^{-1}(\xi_{\phi}^*, \btheta^*) \right\}\right)^{1/q}}
{\left(\det\left\{ \bI^{-1}(\xi, \btheta^*) \right\}\right)^{1/q}},$$
where $ \xi_{\phi}^*$ is a D-optimal design.
For other optimality criteria, the design efficiency
is usually given by (\ref{EffA}).

Notice that   $\xi_{\phi}^*$  may not always be available. In those cases,
the OAD on $S_n$$,\xi_{\phi, N}^*$, can be used to compute a modified design efficiency as
$$\widetilde{\Eff}_{\phi}(\xi_{n,\phi}^*)= 
\frac{\phi\left\{ \bI^{-1}(\xi_{\phi, N}^*, \btheta^*) \right\}}
{\phi\left\{ \bI^{-1}(\xi_{n,\phi}^*, \btheta^*) \right\}}.$$ 
Since $ \phi\left\{ \bI^{-1}(\xi_{\phi, N}^*, \btheta^*) \right\}  
\ge \phi\left\{ \bI^{-1}(\xi_{\phi}^*, \btheta^*) \right\},$
it is possible that $\widetilde{\Eff}_{\phi}(\xi_{n,\phi}^*) > 1$. 
Nevertheless, we can still say that $\xi_{n,\phi}^*$ is highly efficient when 
$\widetilde{\Eff}_{\phi}(\xi_{n,\phi}^*) \ge 0.95$.
Additional comments are given in Section \ref{ssec-properties}.
 
We will introduce our proposal for an effective algorithm to find highly efficient $\xi_{n,\phi}^*$ 
for small $n$ and $\xi_{\phi, N}^*$  in Section \ref{ssec-Algorithm}. 
Then we discuss and explore various properties of the algorithm and $\xi_{n,\phi}^*$ in Section \ref{ssec-properties}.
 
\subsection{Algorithms}\label{ssec-Algorithm}

We develop a general algorithm for finding an approximate design $\xi_{\phi,N}^*$ 
for a given design problem, and then construct an exact design $\xi_{n,\phi}^*$ with
high efficiency. In Algorithm \ref{alg}, we first compute the OAD, $\xi_{\phi,N}^*$ via CVX, and then
$\xi_{n,\phi}^*$ is obtained through a SA with a starting design generated from $\xi_{\phi,N}^*$.
To describe the algorithm clearly, we use a general design problem below to explain the detailed
steps in the algorithm.
The objective function is $\Phi(\xi)=\phi\left\{\bI^{-1}(\xi,\btheta^*) \right)$, where $\xi$ is a design on design space $S \subset \R^p$.

\begin{lstlisting}[caption={CVXSADes for computing OADs and OEDs.}, mathescape=true,label=alg,escapechar=\%]
Input:
  * $S_N=\{{\bf u}_1, \ldots, {\bf u}_N \}$: a set of discrete design space
  * Function $\boldsymbol{f}(\bx,\btheta^*)$ or functions $\lambda(\bx,\btheta^*)$ and $\boldsymbol{f}(\bx)$ to compute the information matrix $\bI(\xi,\btheta^*)$ in %\eqref{eq-info1}% or %\eqref{eq-info2}%
  * $\Phi(\xi)=\phi\left\{\bI^{-1}(\xi,\btheta^*)\right\}$: a loss/objective function 
  * $n$: the number of points in the exact design
  * $M$: the number of times to run the annealing algorithm
  * $T_0,~T_{max},\alpha,K,\delta=10^{-5}$: parameters in the annealing algorithm for initial temperature, minimum temperature, cooling factor, number of iterations for each temperature, and tolerance for a stopping criterion, respectively

------------------  Main steps of the algorithm ------------------------
-----------------------------------------------------------------------
Step 1: Compute an OAD $\xi_{\phi, N}^*$ on $S_N$ as a solution to problem %\eqref{eq-OptSN}% via CVX solver;

Step 2: For $j=1,\dots,M$ do 
  2.1: Get an initial exact design from $\xi_{\phi, N}^*$ and denote it  as $\xi_{n,\phi,j}^{(0)}$. The support points of $ \xi_{n,\phi,j}^{(0)}$ are denoted by $\bx_1^{(0)}, \ldots, \bx_m^{(0)}$, which are the same as  those of  $\xi_{\phi, N}^*$. The  weights $w_1^{(0)}, \ldots, w_m^{(0)}$, at $\bx_1^{(0)}, \ldots, \bx_m^{(0)}$,   respectively, are obtained as follows. Let $n_i^{(0)}$ be the rounded integer from $nw_i^*$, satisfying $\sum_{i=1}^m n_i^{(0)} =n$, and let $w_i^{(0)}=n_i^{(0)}/n$ for $i=1, \ldots, m$. Note that $w_1^*, \ldots, w_m^*$ are the weights at the support points of  $\xi_{\phi, N}^*$;
  2.2: Let $t\leftarrow 1$, $T\leftarrow T_0$, $\ell_1\leftarrow 1$ and $\ell_2\leftarrow 0$;
  2.3: Simulated Annealing: while $T>T_{min}$ and $\abs{\ell_2-\ell_1}>\delta$ do
      2.3.1: Make a small change in $\xi_{n,\phi,j}^{(t-1)}$ to get $\xi_{n,\phi,j}^{(t)}$. It is done by moving a randomly selected design point in $\xi_{n,\phi,j}^{(t-1)}$, say $\tilde\bx$, to a new location in $S$, and the new location is randomly generated in a small hyper-cube centred at $\tilde\bx$;
      2.3.2: $\xi_{n,\phi,j}^{(t)}$ is accepted if $\exp\left\{-(
    \Phi(\xi_{n,\phi,j}^{(t)}) -\Phi(\xi_{n,\phi,j}^{(t-1)}))/T\right\} > u^{(t)}$, where $u^{(t)}\sim \mbox{unif}(0,1)$. If it is accepted, then $\ell_1\leftarrow \Phi(\xi_{n,\phi,j}^{(t-1)})$, $\ell_2\leftarrow \Phi(\xi_{n,\phi,j}^{(t)})$ and $t\leftarrow t+1$;
      2.3.3: Let $T\leftarrow \alpha\cdot T$ after using the same temperature for $K$ times;
  2.4: Let $\xi_{n,\phi,j}^*$ be the last accepted design. Compute $\Phi(\xi_{n,\phi,j}^{*})$ and the modified design efficiency $\widetilde{\Eff}_{\phi}(\xi_{n,\phi,j}^*)\leftarrow \frac{ \Phi(\xi_{\phi, N}^{*})}{ \Phi(\xi_{n,\phi,j}^{*})}$;

Step 3: Select best the design from $\{\xi_{n,\phi,j}^*\}_{j=1}^M$ with the highest efficiency $\widetilde{\Eff}_{\phi}$ and denote it as an OED $\xi_{n,\phi}^*$.
\end{lstlisting}
 
\begin{figure}[ht!]
	\centering
 \includegraphics[width=3.5in]{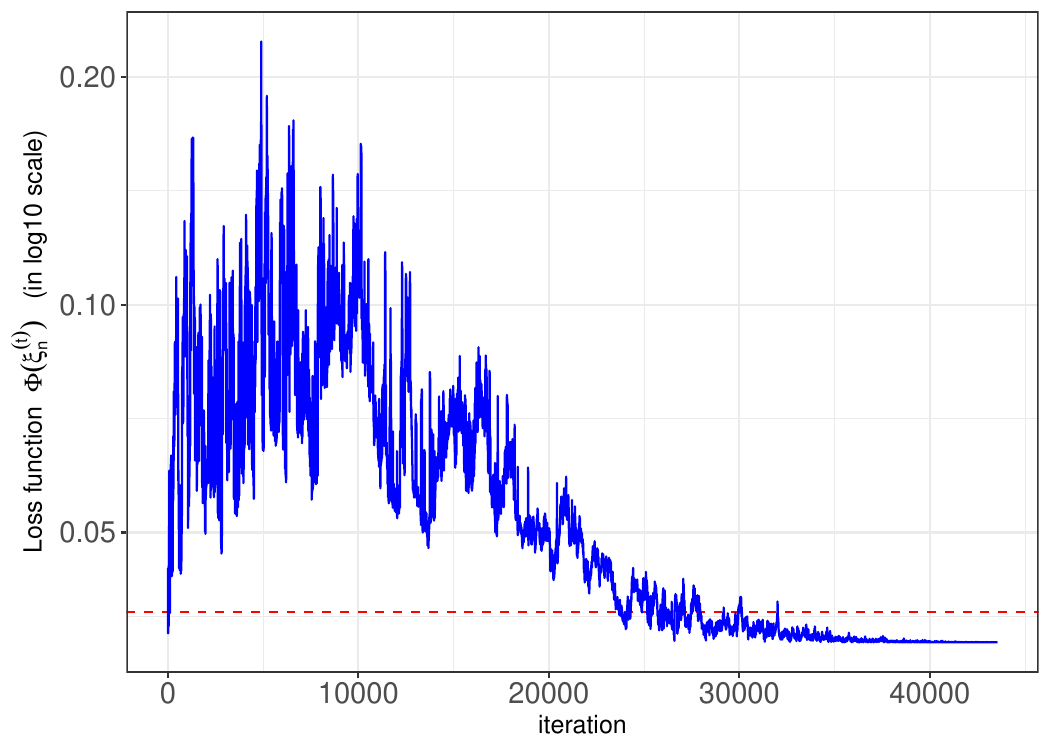}
	\caption{Plot of  loss function values $\Phi(\xi_{n,\phi,j}^{(t)})$ in an annealing algorithm, where the red line is the value of the initial value after rounding $\Phi(\xi_{n,\phi,j}^{(0)})$.}
	\label{fig0}
\end{figure}
\begin{remark}
\citet{wong:2019:cvx,wong:2023:CVX} have provided details for finding OADs via CVX with both single and multiple-objective OADs on $S_N$.
CVX solver can solve problem \eqref{eq-OptSN} easily with 
$N$ as large as 10,000. From our numerical results, $N$ 
does not have to be very large to obtain a highly 
efficient OAD
$\xi_{\phi,N}^*$. Often we can choose $N < 10,000$. In Application \ref{app1-logit} in Section \ref{sec-application}, a
highly efficient OAD
$\xi_{\phi,N}^*$ is obtained with $N=21^2=441$, 
and additional comments are provided there.
\end{remark}

\begin{remark}
To get an initial design $ \xi_{n,\phi}^{(0)}$,
we need to get the rounded integers $n_i^{(0)}$.
When it is not clear how to round $nw_i^*$ for some applications, we can just
take floor or ceiling of $nw_i^*$ such that $\sum_{i=1}^m n_i^{(0)} =n$.
Since $\xi_{n,\phi}^{(0)}$ is just a starting point in the annealing algorithm, the rounding effect is not crucial for the substantial task. As usual, running the annealing algorithm several times is helpful to find
an exact design with a high design efficiency value.
\end{remark}

\begin{remark}
SA algorithm has been used to find optimal designs in the literature, where there are several parameters in the algorithm that need to be adjusted for each optimization problem.  For instance,
\citet{wilmut:2011:two-level-ffd} discussed some strategies for adjusting those parameters. Here,
it is helpful to use a plot of loss function values $\Phi(\xi_{n,\phi,j }^{(t)})$ versus $t$ (iteration)
for checking the annealing parameters, for each fixed $j$. Figure \ref{fig0} (from Application \ref{app2-group} in Section \ref{sec-application}) shows a plot from the annealing algorithm when the parameters are set appropriately.
At the beginning of the search $\Phi(\xi_{n,\phi,j }^{(t)})$ fluctuates, and eventually  as $t$ increases
$\Phi(\xi_{n,\phi.j }^{(t)})$ decreases and converges to a limit.
\end{remark}

\begin{remark}
If an OAD $\xi_{\phi}^*$ on $S$ is available, 
skip Step 1 and replace $\xi_{\phi, N}^*$ by it in Step 2.1.
\end{remark}

\begin{remark}
Since a SA algorithm is used in Step 2.3,
it does not guarantee to find the best exact design.  However,
we can set a required efficiency value, say $0.95$, and run the algorithm  $M$ times to search for a highly efficient exact design. 
\end{remark}

\begin{remark}
    Algorithm \ref{alg} also works for the maximin optimal design by setting
$$\Phi(\xi)= - \min_{1 \le i \le l} ~\Eff_{\phi_i}(\xi).$$ 
\end{remark}

In Section \ref{sec-application}, we provide four representative applications with various models and optimality
criteria to show that Algorithm \ref{alg} is 
effective and works well. 

\subsection{Properties of optimal designs}\label{ssec-properties}

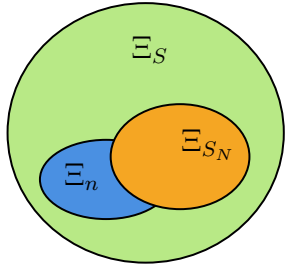
\begin{figure}\label{fig1-tikz}
\centering
\tikzset{every picture/.style={line width=0.75pt}} 

\begin{tikzpicture}[x=0.75pt,y=0.75pt,yscale=-1,xscale=1]

\draw  [fill={rgb, 255:red, 184; green, 233; blue, 134 }  ,fill opacity=1 ] (209,141.5) .. controls (209,105.33) and (240.19,76) .. (278.67,76) .. controls (317.14,76) and (348.33,105.33) .. (348.33,141.5) .. controls (348.33,177.67) and (317.14,207) .. (278.67,207) .. controls (240.19,207) and (209,177.67) .. (209,141.5) -- cycle ;
\draw  [fill={rgb, 255:red, 74; green, 144; blue, 226 }  ,fill opacity=1 ] (225.33,165) .. controls (225.33,153.95) and (240.26,145) .. (258.67,145) .. controls (277.08,145) and (292,153.95) .. (292,165) .. controls (292,176.05) and (277.08,185) .. (258.67,185) .. controls (240.26,185) and (225.33,176.05) .. (225.33,165) -- cycle ;
\draw  [fill={rgb, 255:red, 245; green, 166; blue, 35 }  ,fill opacity=1 ] (261,153.5) .. controls (261,138.86) and (276.67,127) .. (296,127) .. controls (315.33,127) and (331,138.86) .. (331,153.5) .. controls (331,168.14) and (315.33,180) .. (296,180) .. controls (276.67,180) and (261,168.14) .. (261,153.5) -- cycle ;

\draw (270,91.4) node [anchor=north west][inner sep=0.75pt]    {$\mathrm{\Xi} _{S}$};
\draw (295,138.4) node [anchor=north west][inner sep=0.75pt]    {$\mathrm{\Xi}_{S_{N}}$};
\draw (236,155.4) node [anchor=north west][inner sep=0.75pt]    {$\mathrm{\Xi}_{n}$};
\end{tikzpicture}
\caption{Illustration of the three sets of distributions, $\Xi_S$, $\Xi_{S_N}$ and $\Xi_n$ for the optimal designs, OADs and OEDs, respectively.}\label{fig-tikz}
\end{figure}

Figure \ref{fig-tikz} 
 shows the relationship among the three sets of distributions, where $\Xi_n$ is the set 
of exact designs with run size $n$ on $S$. It is obvious
that the optimal designs, 
$\xi_{n,\phi}^*$, 
$\xi_{\phi, N}^*$, and $\xi_{\phi}^*$ for any criterion
on $\Xi_n, \Xi_{S_N},$ and $\Xi_S$, respectively,
satisfy
$$ \phi\left\{ \bI^{-1}(\xi_{n, \phi}^*, \btheta^*)\right\}  \ge
\phi\left\{ \bI^{-1}(\xi_{\phi}^*, \btheta^*)\right\}, ~~ 
\phi\left\{ \bI^{-1}(\xi_{\phi, N}^*, \btheta^*)\right\}  \ge
\phi\left\{ \bI^{-1}(\xi_{\phi}^*, \btheta^*)\right\},$$
for any $n$ and $N$.
 Theorem \ref{thm-1} below shows some asymptotic results 
of $\xi_{n,\phi}^*$ and $\xi_{\phi,N}^*$ under some mild conditions. 

\begin{theorem}\label{thm-1}
Consider a regression model and a design space $S$. Let $\xi_{n,\phi}^*$ be the OED for a given $n$, $\xi_{\phi,N}^*$ be the OAD on $S_N$, and $\xi_\phi^*$ be the OAD design on $S$. Assume all entries of $I(\bx, \btheta^*)$ are continuous functions of
$\bx$ and $S$ is a bounded region,
where $I(\bx, \btheta^*)={\bf f}(\bx, \btheta^*)
{\bf f}^\top(\bx, \btheta^*)$ for model (\ref{M1}),
or $I(\bx, \btheta^*)=\lambda(\bx, \btheta^*)
{\bf f}(\bx){\bf f}^\top(\bx)$ for a GLM. 
In addition, $S_N$ is formed by the Cartesian product of the 
equally spaced points for each design variable. We have the following results:
\begin{enumerate}[(i)]
    \item $\lim_{n \to \infty} \Eff_{\phi}(\xi_{n,\phi}^*) =1$.
    \item  $\lim_{N \to \infty} \Eff_{\phi}(\xi_{\phi, N}^*) =1$.
\end{enumerate}

\end{theorem}

The detailed proof is given in the Appendix. 
The asymptotic results indicate that 
$\xi_{n,\phi}^*$ and $\xi_{\phi,N}^*$
can be highly efficient for large $n$ and $N$.
However, 
in practice we are often interested in exact designs with small $n$, and Algorithm \ref{alg} is very helpful for finding those exact designs.

From Theorem \ref{thm-1},  $\lim_{n \to \infty} \Eff_{\phi}(\xi_{n,\phi}^*) =1$. However,
$\Eff_{\phi}(\xi_{n,\phi}^*)$ may not  be an increasing function of $n$.
For example, if an OAD has 4 support points with equal weights, $1/4, 1/4,
1/4, 1/4$,  then it is easy to construct OEDs for $n$ being  multiples of 4
and those OEDs have $\Eff_{\phi}(\xi_{n,\phi}^*)=1$.
However, when $n$ is not a multiple of 4, it is clear that $\Eff_{\phi}(\xi_{n,\phi}^*)<1$.

When the design space $S$ is discrete, we take $S_N=S$.  Then we have  
$\xi_{\phi}^*=\xi_{\phi, N}^*$.  In this situation, 
efficiency measures $\Eff_{\phi}(\xi_{n,\phi}^*) =
  \widetilde{\Eff}_{\phi}(\xi_{n,\phi}^*)$.
  In general, for any $S$ and $S_N \subset S$, design efficiency measures satisfy
  $$\Eff_{\phi}(\xi_{n,\phi}^*) = \Eff_{\phi}(\xi_{\phi, N}^*) \cdot
\widetilde{\Eff}_{\phi}(\xi_{n,\phi}^*),$$
by their definitions in Sections \ref{ssec-multiple-obj} and \ref{sec-algorithm}. For instance,
if $\Eff_{\phi}(\xi_{\phi, N}^*)=0.97$ and 
$\widetilde{\Eff}_{\phi}(\xi_{n,\phi}^*)=0.98$, then 
$\Eff_{\phi}(\xi_{n,\phi}^*) = 0.9506$.
Since $\widetilde{\Eff}_{\phi}(\xi_{n,\phi}^*)$ 
may be larger than 1, 
$\Eff_{\phi}(\xi_{n,\phi}^*)$ can be larger than 
$\Eff_{\phi}(\xi_{\phi, N}^*)$ for some models and in particular for small $N$.
$\Eff_{\phi}(\xi_{\phi, N}^*)$ may
not be an increasing function of $N$,
but $\lim_{N \to \infty} \Eff_{\phi}(\xi_{\phi, N}^*) =1$
from Theorem \ref{thm-1}.

Algorithm \ref{alg} works well and is effective to find highly efficient $\xi_{n, \phi}^*$ for small or
moderate $n$.
The number of distinct support points in  
$\xi_{n, \phi}^*$ may not be the same as that in $\xi_{\phi}^*$ or $\xi_{\phi, N}^*$.
When $n$ is very large, the SA algorithm in Algorithm \ref{alg} can be slow.
This is true for any SA algorithms.
In this situation,  we can use a rounding method to construct OEDs, as
illustrated in the proof of 
Theorem \ref{thm-1} in the Appendix,
and we can replace $\xi_{\phi}^*$ by $\xi_{\phi, N}^*$ when $\xi_{\phi}^*$
is not available.
In that case the number of distinct support points in  
$\xi_{n, \phi}^*$ are the same as that in $\xi_{\phi}^*$ or $\xi_{\phi, N}^*$.

\section{Applications}\label{sec-application}
We apply our proposed algorithm, CVXSADes, to construct OEDs for various models and various values of $n$. Representative results are given and discussed in four applications below. Application \ref{app1-logit} is for a D-optimality criterion for a logistic model with two design variables, where it is not clear how to round $nw_i$ from OADs.
In addition, we discuss the choice of $N$ and the efficiency of OADs $\xi_{\phi,N}^*$ on $S_N$ as $N$ varies.
Application \ref{app2-group} is about constructing optimal group testing designs over a discrete design space $S$.
The OAD $\xi_{\phi,N}^*$  from Algorithm \ref{alg} is more appropriate than the optimal designs in a paper.
Application \ref{app3-7D} concerns high-dimensional designs, where the design space comprises seven design variables. The high-dimensionality presents a challenge in finding OADs and OEDs. Algorithm \ref{alg} provides an alternative method to find optimal designs and  gives better approximate designs than those in a couple of examples in \citet{xu:2019:design-HD}.
Application \ref{app4-maximin} shows results for maximin optimal designs in which multiple objectives compete against each other. In each of the applications, we demonstrate the usefulness and effectiveness of our proposed algorithm.

\begin{application}{(\textbf{Two-variable logit model})}\label{app1-logit}
Consider a two-variable binary logistic regression model with interaction, as discussed in \citet{haines:2018:two-logistic}, where it was used to study the effectiveness of different combinations of the concentration of two insecticides. The model is given as 
$$\mbox{logit}(p)=\log\left(\frac{p}{1-p} \right) =\theta_0+\theta_1x_1+\theta_2x_2+
\theta_{12}x_1x_2,$$
where  $p$ is the probability of success, i.e., $p=P(Y=1)$, $Y$ is a binary response variable,
$x_1$ and $x_2$ are two design variables, such as the doses of two drugs.
D-OEDs were studied and derived analytically in \citet{haines:2018:two-logistic} for various parameter values $\btheta^*$
and design spaces.
For this GLM, the information matrix of
design $\xi(\bx)$  in \eqref{eq-info2}
becomes
\begin{eqnarray}
\bI(\xi, \btheta^*) =\sum_{i=1}^k w_i 
\frac{\exp(\bff^\top({\bf v}_i)\btheta^*)}{
\left( 1+  \exp(\bff^\top({\bf v}_i)\btheta^*) \right)^2}
\bff({\bf v}_i)
\bff^\top({\bf v}_i),
\label{InfoGLM}
\end{eqnarray}
where $\bff(\bx) =(1,x_1,x_2,x_1x_2)^\top$.

\begin{figure}[ht!]
	\centering
	\includegraphics[width=5.5in]{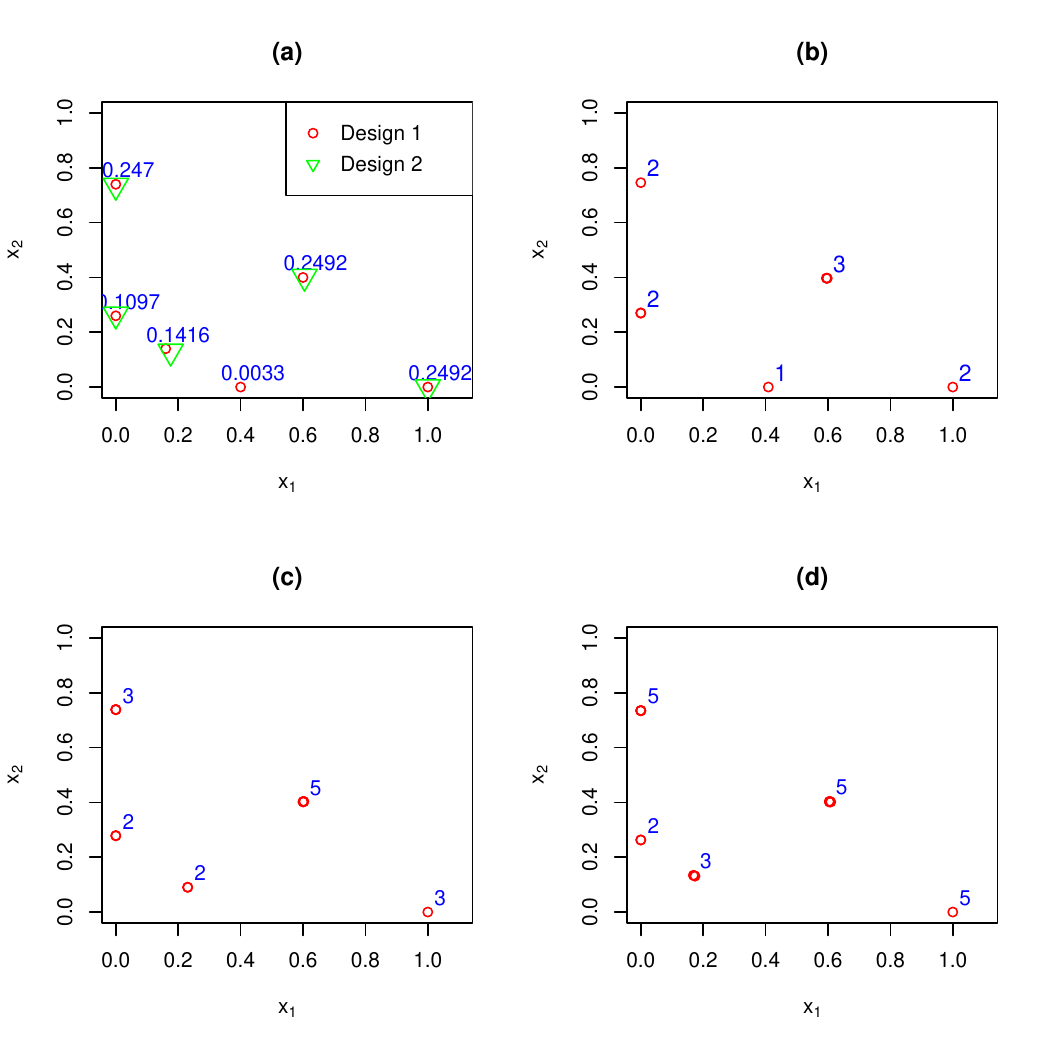}
	\caption{Plots of support points of
	D-OAD and D-OED in Application \ref{app1-logit}, case (i) with parameters $\btheta = (-3,4,6,1)^\top$, design space $S=[0,1]^2$ and $N=51^2$ grid, where the numbers in blue-colour are the weights or numbers of observations.
	\textbf{(a)} Design 1 (D-OAD on $S_N$)  and
	     Design 2 from \citet[Example 4.2(b)]{haines:2018:two-logistic},
	\textbf{(b)} D-OED for $n=10$,
	\textbf{(c)} D-OED for $n=15$,
	\textbf{(d)} D-OED for $n=20$.}
	\label{fig1}
\end{figure}

Since the D-optimality is used as a loss function $\phi$,
corresponding OADs and OEDs are denoted as
D-OADs and D-OEDs.
Using Algorithm \ref{alg} we compute D-OADs $\xi_{\phi, N}^*$
on $S_N$ and construct D-OEDs for various
values of $N$ and $n$. Then we compare $\xi_{\phi, N}^*$ with 
$\xi_{\phi}^*$ in \citet{haines:2018:two-logistic} and comment on the D-OEDs.
We have worked on several cases of $\btheta^*$
and design spaces.  Here we give representative results for two cases:\\
(i) $\btheta^*=(-3, 4, 6, 1)^\top$ and $S=[0,1]^2$ as in 
\citet[Example 4.2(b)]{haines:2018:two-logistic},\\
(ii) $\btheta^*=(-2.2054, 13.5803, 2.2547, 1.6262)^\top$ and $S=[0,2]^2$ as in
\citet[Section 5]{haines:2018:two-logistic}.
 
For case (i), D-OADs and D-OEDs are plotted in Figure \ref{fig1}.
In the approximate design in Figure \ref{fig1}(a),  $S_N$ includes $N=51^2$ grid points in $S$,
which is formed by Cartesian product of 51 equally spaced points in [0,1] in each dimension. 
Note that  $\xi_{\phi, N}^*$ and $\xi_{\phi}^*$ have 5 support points that are almost the same,
but $\xi_{\phi, N}^*$ has one extra support point with a very small weight (0.0033).
The weights of  $\xi_{\phi, N}^*$ are displayed there, and the weights of  $\xi_{\phi}^*$ are slightly 
different and not shown in the plot for clear presentation.
In addition, the loss function values of the two designs are almost the same, with
$\Eff_D(\xi_{N,\phi}^*)=0.9998$.
Three D-OEDs with $n=10, 15,$ and 20 are plotted in 
Figure \ref{fig1}(b), (c) and (d), respectively,
with efficiency 
$\Wtilde{\Eff}_{\phi}(\xi_{n,\phi}^*)=0.9836, 0.9785$ and $1.0001$.
This indicates that these D-OEDs are highly efficient.
The support points of each D-OED are clustered around 5 points.

\begin{figure}[ht!]
	\centering
 \includegraphics[width=5.5in]{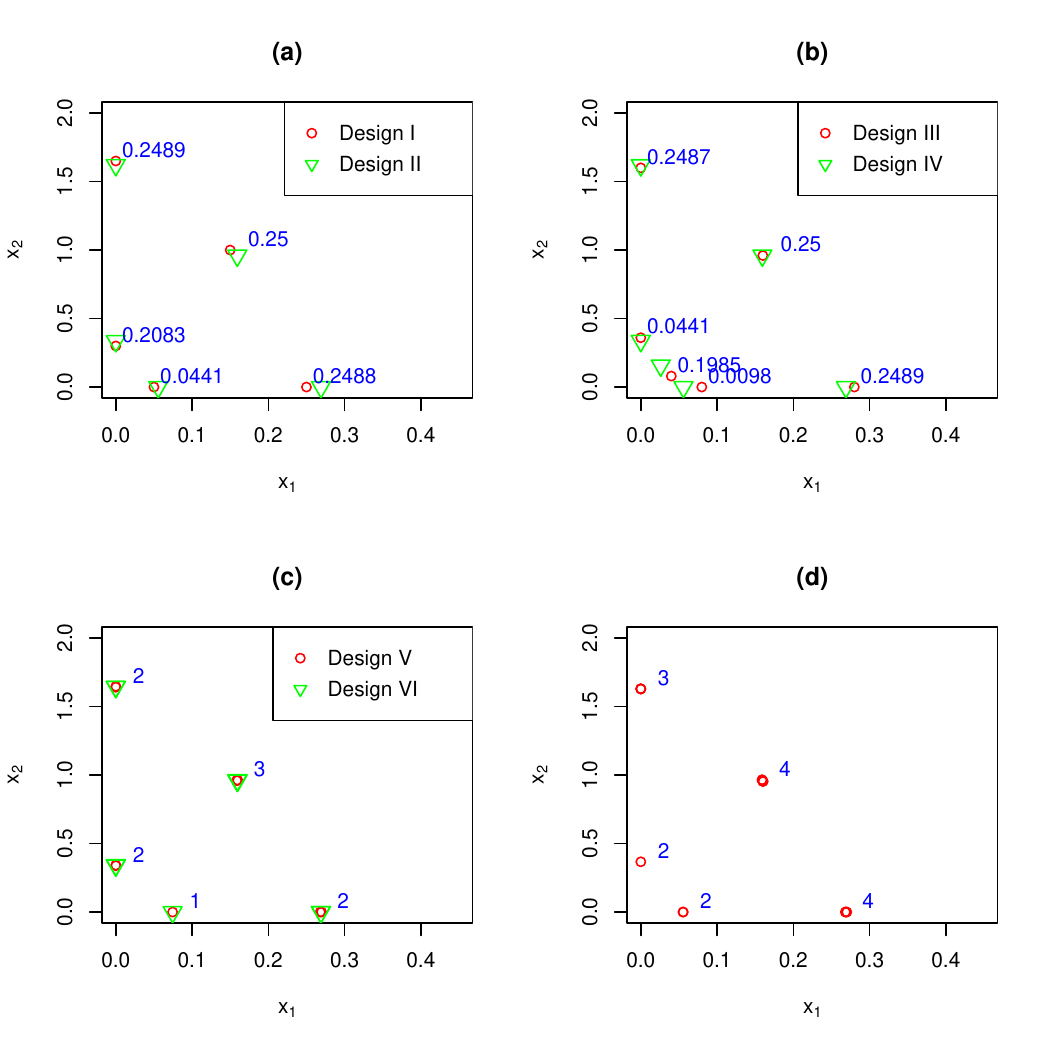}
	\caption{Plots of support points of
	D-OADs and D-OEDs, where the numbers in blue-colour
	are the weights or numbers of observations.
	(a) Design I (D-OAD on $S_N$ with $N=41^2$)  and
	     Design II from \citet[the 5-support-point design in Section 5]{haines:2018:two-logistic},
	(b) Design III (D-OAD on $S_N$ with $N=51^2$)  and
	     Design IV from \citet[the 6-support-point design in Section 5]{haines:2018:two-logistic},
	(c) Design V (D-OED for $n=10$ started from Design I),
	     Design VI (D-OED for $n=10$ started from Design III),
	(d) D-OED for $n=15$ started from Design III.}
	\label{fig2}
\end{figure}

For case (ii),  D-OADs and D-OEDs are plotted in Figure \ref{fig2}.
There are three D-OADs in \citet[Section 5]{haines:2018:two-logistic},
which have 4, 5, and 6 support points, respectively.
The one with 6 support points has the smallest value of ${\phi}(\xi_{\phi}^*)$, and
we use it for the efficiency calculation of $\xi_{\phi, N}^*$ and $\xi_{n, \phi}^*$.
Several values of $N$ are used to compute $\xi_{\phi, N}^*$ and $\xi_{n, \phi}^*$.
Representative results are given in Table \ref{table2}, which indicate that 
${\Eff}_{\phi}(\xi_{\phi, N}^*)$ increases as $N$ increases, 
${\Eff}_{\phi}(\xi_{n,\phi}^*)$ seems to be fluctuated a bit, and 
but both $\xi_{n, \phi}^*$ and $\xi_{\phi, N}^*$ have a high efficiency with the $N$ in the latter being as small as $21^2$.
The interpretation of having a constant ${\Eff}_{\phi}(\xi_{n,\phi}^*)$
is that the annealing algorithm can effectively find highly efficient exact designs starting from
different $\xi_{\phi, N}^*$.
This implies that the choice of $N$ does not affect the OED much, and in practice, we can choose a moderate $N$. In Figure \ref{fig2} (a) and (b), $\xi_{\phi, N}^*$ and 
$\xi_{\phi}^*$ are plotted.  For clear presentation, the plots only display the weights in  
$\xi_{\phi, N}^*$, which can also have 5 and 6 support points with different values of $N$.
The locations of the support points in $\xi_{\phi, N}^*$ are similar to those in $\xi_{\phi}^*$.
In Figure \ref{fig2}(c) and (d), D-OEDs are plotted.  Design V and VI are the same,
but the annealing algorithm used different starting designs.
In \citet[Section 5]{haines:2018:two-logistic}, it is not clear how to get D-OEDs from 
$\xi_{\phi}^*$ with any given value of $n$.  Algorithm \ref{alg} provides an effective way to construct
D-OEDs for any $n$. The exact design in Figure \ref{fig2}(d)
has ${\Eff}_{\phi}(\xi_{n,\phi}^*)=0.9935$ for $n=15$.

\begin{table}[ht!]
\renewcommand{\arraystretch}{1} 
\linespread{1}\selectfont\centering
\caption{Efficiency  of $\xi_{\phi, N}^*$ and $\xi_{n, \phi}^*$ (with $n=10$)
for case (ii) in Application \ref{app1-logit}.}
\begin{center}
\begin{tabular}{crr} \hline
$N$  & ~~~${\Eff}_{\phi}(\xi_{\phi, N}^*)$  &~~~${\Eff}_{\phi}(\xi_{n,\phi}^*)$  \\ \hline
$21^2$  &0.9716 &0.9822 \\
$31^2$  &0.9901 & 0.9513\\
$41^2$  &0.9961 & 0.9793\\
$51^2$  & 0.9985& 0.9794\\
$81^2$  & 0.9984& 0.9822 \\ \hline
\end{tabular}
\end{center}
\label{table2}
\end{table}

\end{application}

\begin{application}{(\textbf{Group testing design for disease prevalence})}\label{app2-group}
Group testing is employed to study rare diseases when testing individuals for a trait is costly \citep{hughes:1994:group,hughes:2000:group}. Instead of taking samples from each individual and testing them individually, it is more cost-efficient to conduct group testing, wherein samples from individuals are pooled as a group and tested together as a unit. In \citet{huang:2017:design-group},
optimal group testing designs are studied, where optimal group sizes are selected for 
group testing experiments.  Since the design space $S$ only includes integer values, finding OADs on discrete design spaces $S_N (=S)$ via CVX is extremely useful.
We demonstrate this for D- and c-optimality criteria in \citet{huang:2017:design-group}, and present
OADs and OEDs for various values of $n$.  In addition, we find and comment on
interesting features in OEDs. 

To present the optimality criteria clearly, we rewrite the information matrix from \citet{huang:2017:design-group} as follows,
\begin{eqnarray}
\bI(\xi, \btheta) =\sum_{i=1}^k w_i \lambda(x_i) \bff(x_i) \bff^\top(x_i),
\label{InfoGroup}
\end{eqnarray}
 where $\btheta=(p_0,p_1,p_2)^\top$,
 $\lambda(x)=1/(\pi(x)(1-\pi(x)))$ with $\pi(x)=p_1-(p_1+p_2-1)(1-p_0)^x$, and
 $$\bff(x)=\left( \begin{array}{c}
 x(p_1+p_2-1)(1-p_0)^{x-1} \\
 1-(1-p_0)^x \\
 -(1-p_0)^x 
 \end{array}    \right).$$
 See the details of the group testing model and 
 information matrix in \citet{huang:2017:design-group}.
 We consider D-optimality and c-optimality
with vector ${\bf c}_1=(1,0,0)^\top$.  We minimize
$\phi\{ \bI^{-1}(\xi, \btheta^*)\}={\bf c}_1^\top \bI^{-1}(\xi, \btheta^*) {\bf c}_1$
to obtain c-optimal designs.
Representative results are given in Table \ref{table3} for $ \btheta^*=(0.07, 0.93, 0.96)^\top$ and
$S_N=S=\{1, 2, 3, \ldots, 61\}$. From \citet{huang:2017:design-group}, they found that the D-OAD
has three support points:  1, 16.79, 61,  which are similar to those of $\xi_{\phi,N}^*$
in Table \ref{table3}.  However,
one of their support points is not an integer, which is not the design space.  Thus,
$\xi_{\phi,N}^*$ is more appropriate for this experiment.
Since $S$ only includes integers, we modify the annealing algorithm by adding 1 or $-1$ to
the selected design point to obtain a new design point in $S$. As usual, we make sure that all new
points are in $S$.
The D-OEDs for various values of $n$ from Algorithm \ref{alg} are the same as those
by rounding $nw_i$ to integers, and they are highly efficient. 
 
For c-optimal designs, $\xi_{\phi,N}^*$ in Table  \ref{table3}
is also more appropriate for this experiment,
since the c-optimal design in \citet{huang:2017:design-group} also includes a non-integer support point (15.68).
It is interesting to notice that the c-OEDs do not have the sample support 
points as in $\xi_{\phi,N}^*$, and the design efficiencies are very high.
One of the support points in  $\xi_{\phi,N}^*$ is $16$ with a large weight 0.6279,
but some of the exact designs include a support point $15$ or $17$ and do not include $16$ as a support point. See the results for $n=10$, $11$  and $14$.  Figure \ref{fig0} gives a plot of loss function value versus iteration number in the annealing algorithm for finding c-OED with $n=12$ where the y-axis is in log $10$ scale. Note that it is easy to find OADs and OEDs using Algorithm \ref{alg} for any 
$ \btheta^*$ value, any integer design space, and any optimality criterion. 

\begin{table}[ht!]
\renewcommand{\arraystretch}{1}
\linespread{1}\selectfont\centering
\caption{c- and D-OADs, and c-, D-OEDs, $\xi_{\phi, N}^*$ and $\xi_{n, \phi}^*$ in 
Application \ref{app2-group}, where Per($i, j, k$) means any permutation of $i, j, k$.}
\begin{center}
\begin{tabular}{crrrr} \hline
D-optimality  &  support points  & weights/observations & $ {\phi}(\bI^{-1})$  & 
$\widetilde{\Eff}(\xi_{n, \phi}^*)$ \\ \hline
approximate $\xi_{\phi,N}^*$ & 1,  17,  61& 1/3, 1/3, 1/3& 0.1448& 1.0000\\
$n=10$  & 1,  17,  61& Per(4,3,3)&0.1462& 0.9906\\
$n=11$  &1,  17,  61 &Per(4,4,3) &0.1461&0.9912\\
$n=12$  & 1,  17,  61&4, 4, 4&0.1448&1.0000\\
$n=13$  & 1,  17,  61&Per(5,4,4)&0.1457& 0.9944 \\
$n=14$  & 1,  17,  61&Per(5,5,4)&0.1456& 0.9946\\ \hline
c-optimality  &  support points  & weights/observations & $ {\phi}(\bI^{-1})$  & 
$\widetilde{\Eff}(\xi_{n, \phi}^*)$ \\ \hline
approximate $\xi_{\phi,N}^*$ & 1, 16, 61&0.1310,  0.6279,  0.2411& 0.0354& 1.0000\\
$n=10$  &1,    17,    61 &1,     6,    3&0.0361&0.9799\\
$n=11$  &1,    17,    61 &1,     7,     3&0.0361&0.9808\\
$n=12$  &1,    15,    16,    61 &2,     4,     3,     3&0.0358&0.9891\\
$n=13$  &1,    15,    16,    61 &2,     7,     1,     3&0.0355&0.9968  \\
$n=14$  &1,    15,    61 &2,     9,     3&0.0355&0.9970 \\ \hline
\end{tabular}
\end{center}
\label{table3}
\end{table} 
\end{application}

\begin{application}{(\textbf{Seven-dimensional design space})}\label{app3-7D}
In real-world applications, experiments often involve many design variables, and efficient designs can help save resources and prevent wasted time.
Consider a logistic model with seven design variables and its information matrix
is similar to that in \eqref{InfoGLM} with
$\bff(\bx)=(1,x_1,x_2,x_3,x_4,x_5,x_6,x_7)^\top$.
\citet{xu:2019:design-HD} proposed an innovative algorithm using differential evolution for finding D-OADs with several design variables, and used this model as an example. When the design space $S=[-1, 1]^7$, the D-OAD in \citet[Table 4]{xu:2019:design-HD}
with 
\[
\btheta^*=(-0.4926,-0.6280,-0.3283,0.4378,0.5283,-0.6120,
 -0.6837,-0.2061)^\top
\]
has 48 support points.
The corresponding weights for the 48 support points are,
 0.0230,  0.0160,  0.0255, \ldots,  0.0269.   
However, it is challenging to implement the D-OAD in practice.  
For instance, given a run size, let us say $n=30$, how we construct the exact design remains unclear.
We cannot have all the 48 support points in the
exact design. In addition, it is not easy to round
$nw_i$ as $nw_i < 1$ for many of the support points in the D-OAD.

Algorithm \ref{alg} can be used construct D-OADs and D-OEDs
for various design spaces
and $n$.  
Representative results are given in Tables \ref{table4} and \ref{table5}.
For each design variable, we take 4 equally spaced points on $[-1, 1]$, i.e., $
-1, -1/3, 1/3, 1$, and construct $S_N$ by including the Cartesian product of the 
equally spaced points for the 7 variables, which gives  
$N=4^7=16,384$.  Table \ref{table4} presents the  D-OAD $\xi_{\phi, N}^*$ via CVX.  It is clear that $\xi_{\phi, N}^*$ has 29 support points, which is much less than 48,
the number of support points in \citet{xu:2019:design-HD}.  In addition,
$\xi_{\phi, N}^*$ has  a loss function value
$\left(
\det \left\{ \bI^{-1}(\xi_{\phi, N}^*, \btheta^*) \right\}\right)^{1/q}
=4.9485$, which is smaller than $4.9573$, the loss function value 
of the approximate design in \citet{xu:2019:design-HD}.
Thus, CVX solver finds a better  D-OAD with a smaller loss function and
a smaller number of support points.
Table \ref{table5} gives an exact design $\xi_{n, \phi}^*$ with $n=30$, which has 22 support points.
Some of the support points are not at the corners of the hyper-cube $[-1, 1]^7$,
since the annealing algorithm allows us to make small changes of design points in $S$.
This $\xi_{n, \phi}^*$ has a loss function value 
$\left(
\det \left\{ \bI^{-1}(\xi_{n,\phi}^*, \btheta^*) \right\}\right)^{1/q}
=5.1231$, which yields an efficiency
$\widetilde{\Eff}(\xi_{n, \phi}^*) = 0.9659$. 



 
\begin{table}[ht!]
\renewcommand{\arraystretch}{1}
\linespread{1}\selectfont\centering
\caption{D-OAD $\xi_{\phi, N}^*$ via CVX with $N=4^7$.}
\begin{center}
\begin{small}
\begin{tabular}{crrrrrrrr} \hline
Support point & $x_1$ & $x_2$ & $x_3$ & $x_4$ & $x_5$ & $x_6$ & $x_7$ &  weight \\ \hline
 1  &  -1 &  -1  & -1 &  -1 &  -1  &  1 &   1&0.0627 \\
 2  &  -1 &  -1   &-1  &  1  & -1  &  1  & -1&0.0732\\
 3   & -1  & -1  & -1  &  1  &  1  & -1  &  1&0.0487\\
 4  &  -1  & -1  &  1  & -1  &  1  & -1  & -1&0.0499\\
 5 &   -1  & -1  &  1  & -1  &  1  &  1  & -1&0.0460\\
 6  &  -1  &  1  & -1 &  -1  & -1   &-1  & -1&0.0088\\
 7 &   -1  &  1  & -1 &  -1  &  1  & -1  & -1&    0.0561\\
 8  &  -1   & 1  & -1  &  1  & -1   & 1  &  1&0.0212\\
 9  &  -1   & 1  & -1  &  1   & 1  & -1   &-1&0.0226\\
 10 &  -1   & 1 &   1  & -1  & -1  & -1  &  1&0.0840\\
 11 &  -1   & 1  &  1 &   1  & -1  &  1  & -1&0.0306\\
 12 &  -1   & 1   & 1  &  1  &  1   &-1  &  1&0.0023\\
 13  & -1  &  1   & 1   & 1  &  1   & 1  &  1&0.0730\\
 14 &   1  & -1   &-1  & -1  & -1 &  -1  &  1&0.0135\\
 15 &   1  & -1 &  -1  &  1  & -1  & -1  &  1&    0.0217\\
 16  &  1  & -1  & -1  &  1  &  1  & -1 &   1&0.0415\\
 17 &   1 &  -1  &  1  & -1  & -1  & -1  &  1&0.0409\\
 18  &  1  & -1  &  1  & -1  & -1  &  1 &  -1&0.0375\\
 19  &  1  & -1  &  1  & -1  &  1  & -1  & -1&0.0073\\
 20  &  1  & -1   & 1   & 1  & -1   & 1  &  1&0.0255\\
 21  &  1  & -1  &  1  &  1  &  1  & -1  & -1&0.0489\\
 22 &   1  & -1  &  1  &  1 &   1  &  1  & -1&0.0100\\
 23 &   1  &  1  & -1   &-1  & -1   &-1  & -1&0.0491\\
 24  &  1  &  1  & -1   & 1  & -1  & -1 &  -1&0.0404\\
 25  &  1  &  1 &   1   &-1  & -1   &-1  &  1&0.0042\\
 26  &  1  &  1  &  1  &  1   &-1   &-1   & 1&0.0058\\
 27  &  1  &  1  &  1   & 1  & -1   & 1   &-1&0.0420\\
 28  &  1  &  1   & 1   & 1  & -1  &  1  &  1&0.0033\\
 29  &  1  &  1  &  1   & 1  &  1   &-1 &   1&0.0295
\\ \hline
\end{tabular}
\end{small}
\end{center}
\label{table4}
\end{table} 

\begin{table}[ht!]
\renewcommand{\arraystretch}{1}
\linespread{1}\selectfont\centering
\caption{A D-OED $\xi_{n, \phi}^*$  with $n=30$, where $n_i$ denotes the number of observations at each support point.}
\begin{center}
\begin{small}
\begin{tabular}{crrrrrrrr} \hline
Support point & $x_1$ & $x_2$ & $x_3$ & $x_4$ & $x_5$ & $x_6$ & $x_7$ &  $n_i$ \\ \hline
  1& -1     & -1     & -1     & -1     & -1     & 1      & 1      & 1      \\
   2&     -1     & -1     & -1     & 1      & -1     & 1      & -1     & 2      \\
   3&     -1     & -1     & -1     & 1      & -1     & 1      & -0.9999 & 1      \\
   4&     -1     & -1     & -1     & 1      & 1      & -1     & 1      & 2      \\
    5&    -1     & -1     & 1      & -1     & 1      & -1     & -1     & 2      \\
    6&    -1     & -1     & 1      & -1     & 1      & 1      & -1     & 2      \\
    7&    -1     & 1      & -1     & -1     & -1     & -1     & -1     & 1      \\
    8&    -1     & 1      & -1     & -1     & 1      & -1     & -1     & 2      \\
    9&    -1     & 1      & -1     & 1      & -1     & 1      & 1      & 1      \\
    10&    -1     & 1      & -1     & 1      & 1      & -1     & -1     & 1      \\
     11&   -1     & 1      & 1      & -1     & -1     & -1     & 1      & 3      \\
     12&   -1     & 1      & 1      & 1      & -1     & 1      & -1     & 1      \\
      13&  -1     & 1      & 1      & 1      & 1      & 1      & 1      & 2      \\
      14&  -0.9996& 1      & 1      & 1      & 1      & -1     & 1      & 1      \\
      15&  -0.9995& -1     & -1     & -1     & -1     & 0.9995 & 1      & 2      \\
      16&  1      & -1     & -1     & 1      & 1      & -1     & 1      & 1      \\
      17&  1      & -1     & 1      & -1     & -1     & -1     & 1      & 1      \\
      18&  1      & -1     & 1      & -1     & -1     & 1      & -1     & 1      \\
      19&  1      & -1     & 1      & 1      & 1      & -1     & -1     & 1      \\
      20&  1      & 0.9998 & -1     & -1     & -1     & -1     & -1     & 1      \\
      21&  1      & 1      & -1     & 1      & -1     & -1     & -1     & 1      \\
      22&  1      & 1      & 1      & 1      & -1     & 1      & -0.9978& 1      \\
 \hline
\end{tabular}
\end{small}
\end{center}
\label{table5}
\end{table} 
\end{application}

\begin{application}{(\textbf{Maximin design for competing criteria in dose finding study})}\label{app4-maximin} In the dose-finding study, the aim is to find a model that characterizes the dose-response relationship effectively \citep{dette:2008:dose-finding}. However, a complex decision-making process is required due to its complexity and other external considerations, such as efficacy and ethics tradeoffs. Often, more than one response model is necessary, and determining how to maximize the efficiency of the design for a model becomes a practical question. 
Here, we consider maximin optimality criteria for multiple objectives and construct maximin exact designs. We use one application with four response models
in \citet{wong:2023:CVX}, which is also studied in \citet{bretz:2010:practical}.
There are one linear response model, two Emax models with different true parameter values,
and one logistic model.
Let $\bI_i(\xi, \btheta_i^*)$, $i=1, \ldots, 4$, be the information matrices for the four models,
respectively, where $\btheta_i^*$ is the true parameter for model $i$, the same as in \citet{wong:2023:CVX}.

We define maximin optimal designs in the same way as in Section \ref{ssec-multiple-obj}.
If $\phi_i(\xi)=\tr\left\{ \bI_i^{-1}(\xi, \btheta_i^*)\right\}$ for $i=1, \ldots, 4$,
we call it as maximin A-optimal design (maximin A-OAD/A-OED).
If $\phi_i(\xi)=\det\left\{ \bI_i^{-1}(\xi, \btheta_i^*)\right\}$ for $i=1, \ldots, 4$,
we call it as maximin D-optimal design (maximin D-OAD/D-OED).

\begin{figure}[ht!]
	\centering
	\includegraphics[width=5.5in]{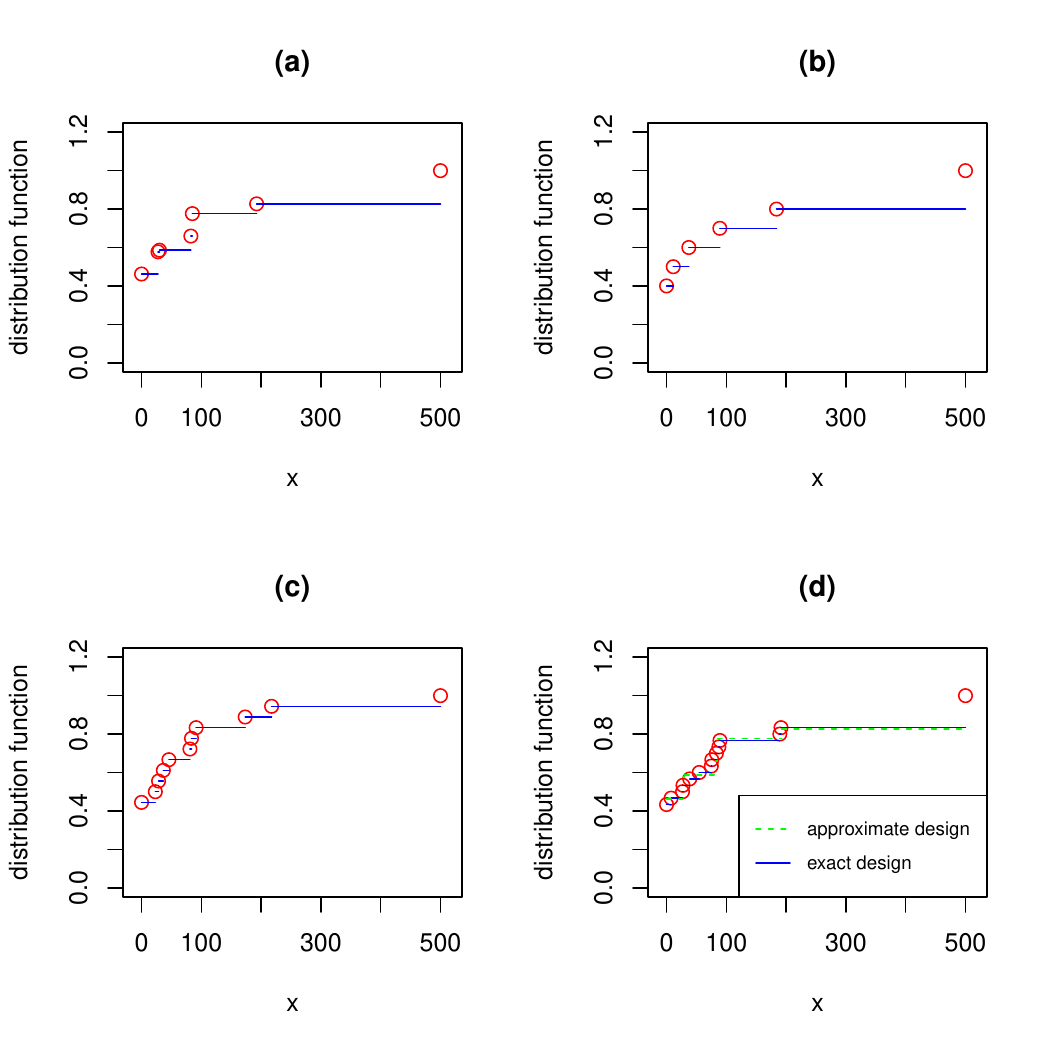}
	\caption{Plots of distribution functions of maximin A-optimal designs.
	(a) approximate design,
	(b)  exact design for $n=10$,
	(c) exact design for $n=20$,
	(d) exact design for $n=30$ and approximate design. }
	\label{fig3}
\end{figure}

\citet{wong:2023:CVX} discussed an algorithm to compute maximin OADs on $S_N$, and the main idea is to transform the maximin problem
into a convex optimization problem and use CVX to find solutions.
Applying the algorithm in \citet{wong:2023:CVX}, we can easily obtain
maximin A- and D-OADs $\xi_{\phi,N}^*$ via CVX in Step 1 of Algorithm \ref{alg}. Then we can compute maximin 
A- and D-OEDs for various values of $n$.
Representative results are plotted in Figures \ref{fig3} and \ref{fig4}, where $S=[0, 500]$,
$N=201$, and $n=10$, 20, and 30 are used.
In Figure \ref{fig3}, the distribution functions of maximin 
A-OADs and maximin A-OEDs are plotted.  From  Figure \ref{fig3}(d), the maximin 
A-OAD and maximin A-OED with $n=30$ are very similar. The maximin A-OAD has $\min \{ \widetilde{\Eff}_{\phi_1}(\xi), \ldots, \widetilde{\Eff}_{\phi_4}(\xi)\}
=0.7155$, and the maximin A-OEDs have $\min \{ \widetilde{\Eff}_{\phi_1}(\xi), \ldots, \widetilde{\Eff}_{\phi_4}(\xi)\}=
0.6813$, $0.6983$ and $0.7121$
for $n=10$, $20$ and $30$, respectively.

In Figure \ref{fig4}, the distribution functions of maximin D-OAD and D-OED are plotted. The maximin D-OED has 5 support points, while the exact ones have more than 5 support points. The maximin D-OAD has $\min \{ \widetilde{\Eff}_{\phi_1}(\xi), \ldots, \widetilde{\Eff}_{\phi_4}(\xi)\} =0.8538$, and the maximin D-OEDs have 
$\min \{ \widetilde{\Eff}_{\phi_1}(\xi), \ldots, \widetilde{\Eff}_{\phi_4}(\xi)\}=$ 
$0.8371$, $0.8420$ and $0.8459$ 
 for $n=10$, 20 and 30, respectively.  From our numerical results,
$\min \{ \widetilde{\Eff}_{\phi_1}(\xi), \ldots, \widetilde{\Eff}_{\phi_4}(\xi)\}$
of the maximin D-OED or A-OED increases as $n$ increases.

\begin{figure}[ht!]
	\centering
    \includegraphics[width=5.5in]{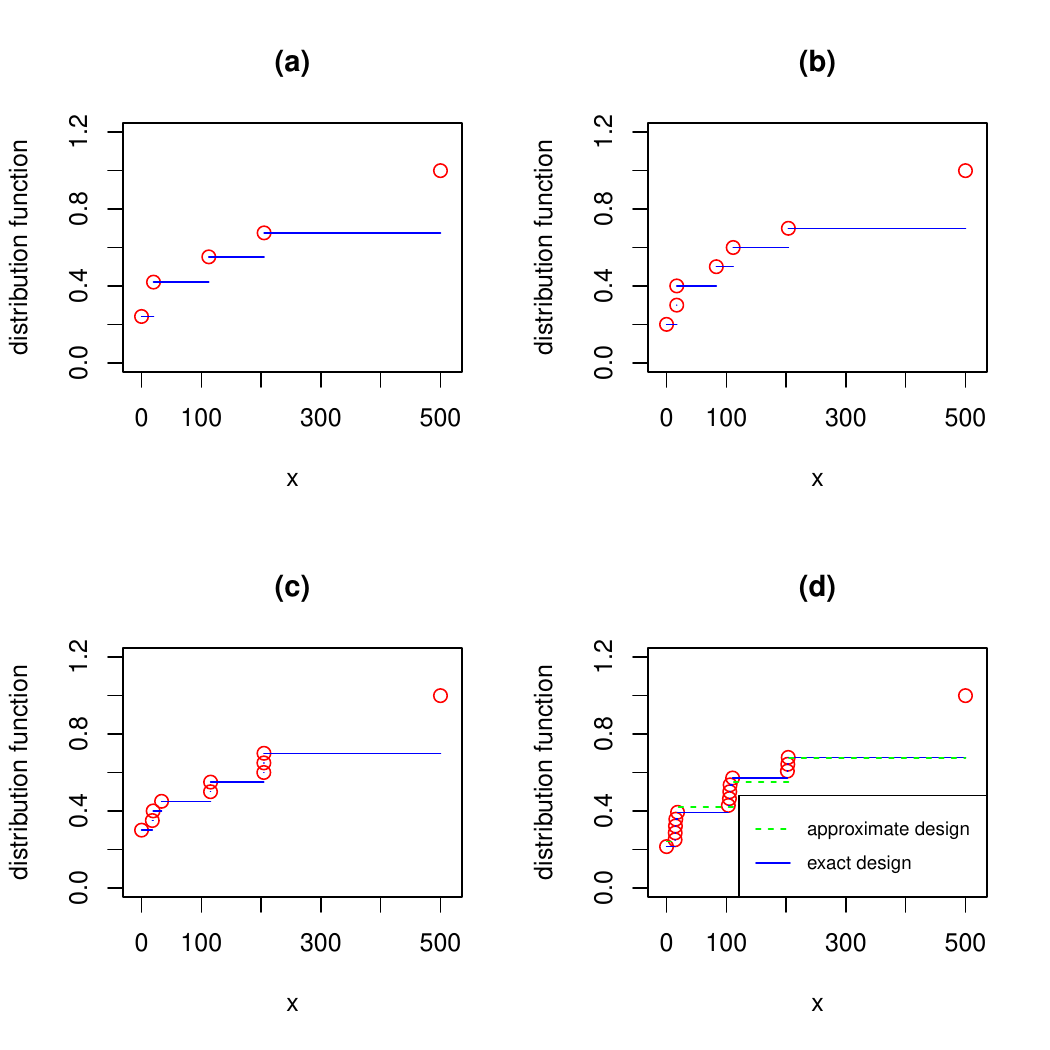}
	\caption{Plots of distribution functions of maximin D-optimal designs.
	(a) approximate design,
	(b) exact design for $n=10$,
	(c) exact design for $n=20$,
	(d) exact design for $n=30$ and approximate design. }
	\label{fig4}
\end{figure}

The gap between the distribution functions of the approximate and exact designs
reflects the differences between their weights and slightly different support points.
For an exact design with $n$ runs, 
the weights have to be 
multiples of $1/n$. 
There are usually more support points in exact designs than those in approximate designs.
Table \ref{table6} presents a maximin A-OED with $n=20$. There are 
11 support points.  Algorithm \ref{alg} is flexible to find highly efficient maximin OEDs with various values of $n$.  

\begin{table}[ht!]
\centering
\renewcommand{\arraystretch}{1}
\linespread{1}\selectfont\centering
\caption{A maximin A-OED with $n=20$, where $n_i$ denotes the number of observations at each support point.}
\resizebox{1\textwidth}{!}{
\begin{small}
\begin{tabular}{c|ccccccccccc}
\hline
\textbf{support point} & 0 & 23.07 & 28.47 & 36.48 & 45.77 & 80.95 & 83.54 & 91.30 & 173.32 & 217.51 & 500 \\
\textbf{$n_i$} & 8 & 1 & 1 & 1 & 1 & 1 & 1 & 1 & 1 & 1 & 3 \\
\hline
\end{tabular}
\end{small}
}
\label{table6}
\end{table}
\end{application}

\section{Conclusion}
\label{sec-conclusion}
Constructing an OED is challenging, as it involves solving an integer programming problem, which is NP-complete. The conventional methods for obtaining the exact design either involve deriving analytical solutions for low-dimensional problems or rounding approximate designs. However, closed-form solutions usually do not exist for designs involving more than two variables, and rounding may not yield exact designs with high efficiency. We have developed a general algorithm to search for OAEs and highly efficient OEDs. This algorithm is applicable to any criterion with a convex loss function, any design spaces, and any sample sizes $n$, although it is particularly useful when $n$ is small or moderate. For very large $n$, a rounding method still performs well in obtaining highly efficient OEDs. Notably, our algorithm also computes the OADs concurrently with OEDs. While there are other numerical algorithms for finding exact designs, they do not compute approximate designs, so it is difficult to access the efficiency of the resulting exact designs; some numerical methods are provided in Section \ref{sec-introduction}. By computing the OADs, we can assess the efficiency of the OEDs relative to the OADs through the modified design efficiency measure. Our method offers a new and alternative approach to finding highly efficient exact designs.

We have provided four applications to demonstrate the effectiveness of our algorithm, which include single and multiple-objective optimality criteria, discrete and high-dimensional design
spaces, linear, nonlinear and GLMs.
Our proposed algorithm can find highly efficient exact designs for small $n$ quickly, which is very useful for practical applications. Full implementation of the proposed algorithm is available online for practitioners to use in other real-world applications.

\section*{Appendix: Proofs and derivations}\label{sec-appendix}

\noindent{\bf Proof of Theorem \ref{thm-1}:} 
Write the information matrices of design $\xi(\bx)$ in \eqref{eq-info1}
and \eqref{eq-info2} using  a general form as
\begin{eqnarray}
\bI(\xi, \btheta^*) = \sum_{i=1}^k w_i I({\bf v}_i, \btheta^*),
\label{InfoG}
\end{eqnarray} 
where $I(\bx, \btheta^*)=
\bff(\bx,\btheta^*) \bff^\top(\bx,\btheta^*)$ in
\eqref{eq-info1}, or $\lambda(\bx, \btheta^*) \bff(\bx) \bff^\top(\bx)$ in
\eqref{eq-info2}.
Suppose the OAD $\xi_{\phi}^*$ has support points $\bx_1^*, \ldots, \bx_m^* \in S$
with corresponding weights, $w_1^*, \ldots, w_m^*$, respectively.

For part (i), we want to show that  
$\lim_{n \to \infty} \Eff_{\phi}(\xi_{n,\phi}^*) =1$.

For each $n$, compute $a_i=n \cdot w_i^*$ for $i=1, \ldots, m$. 
For large $n$ such that $a_i \ge 1$ for all $i=1, \ldots, m$, 
we construct an exact design $\xi^{(n)}$ as follows.  
The support points of $\xi^{(n)}$ are the same as those of $\xi_{\phi}^*$.
We choose integer
$n_i$ to be either the floor or the ceiling of $a_i$ and satisfy $\sum_{i=1}^m n_i=n$.
Since   $a_i \ge 1$ for all $i=1, \ldots, m$, it is clear that  $n_i \ge 1$
and $|n_i -a_i| \le 1$ for all $i=1, \ldots, m$.
Define the weights of exact design $\xi^{(n)}$  to be 
$n_1/n, \ldots, n_m/n$ for 
support points $\bx_1^*, \ldots, \bx_m^*$, respectively.

Let $\delta_i =a_i - n_i$ for $i=1, \ldots, m$. Then 
$w_i^* - n_i/n =w_i^* - a_i/n +\delta_i/n=\delta_i/n$. Note that $|\delta_i| \le 1$.
Evaluate the information matrix of the exact design $\xi^{(n)}$ using \eqref{InfoG},
\begin{eqnarray*}
\bI(\xi^{(n)}, \btheta^*) &=& \sum_{i=1}^m \frac{n_i}{n} 
~I(\bx_i^*, \btheta^*) \\
&=& \sum_{i=1}^m \left( w_i^* - \delta_i/n \right) I(\bx_i^*, \btheta^*) \\
&=& \bI(\xi_{\phi}^{*}, \btheta^*)  - 
\frac{1}{n} \sum_{i=1}^m \delta_i I(\bx_i^*, \btheta^*) \\
&=&\bI(\xi_{\phi}^{*}, \btheta^*)  - \frac{1}{n} {\bf B},
\end{eqnarray*} 
where ${\bf B} = \sum_{i=1}^m \delta_i I(\bx_i^*, \btheta^*)$.
Since  all entries of $I(\bx, \btheta^*)$ are continuous functions of
$\bx$, $S$ is a bounded region, $m$ is a fixed number, and $|\delta_i| \le 1$,
all entries of ${\bf B}$ are bounded.  Thus,
$\frac{1}{n} {\bf B} \to {\bf 0}$ (a zero matrix), as $n \to \infty$.
This implies that $\bI(\xi^{(n)}, \btheta^*) \to 
\bI(\xi_{\phi}^{*}, \btheta^*)$ as $n \to \infty$, which leads to 
$\phi \left\{   \bI^{-1}(\xi^{(n)}, \btheta^*) \right\}  \to
\phi \left\{   \bI^{-1}(\xi_{\phi}^{*}, \btheta^*)\right\}$
for commonly used optimality criterion.

Notice that $\xi^{(n)}$ is an OED with $n$ points.  It is clear
that for any $n$ 
$$\phi \left\{   \bI^{-1}(\xi^{(n)}, \btheta^*) \right\}  \ge
\phi \left\{  \bI^{-1}(\xi_{n,\phi}^*, \btheta^*) \right\} 
\ge \phi \left\{  \bI^{-1}(\xi_{\phi}^{*}, \btheta^*)\right\}.$$
From above analysis, 
$\phi \left\{  \bI^{-1}(\xi^{(n)}, \btheta^*) \right\}  \to
\phi \left\{   \bI^{-1}(\xi_{\phi}^{*}, \btheta^*)\right\}$ as 
$n \to \infty$.  Therefore, we must have
$\phi \left\{   \bI^{-1}(\xi_{n,\phi}^*, \btheta^*) \right\}   \to
\phi \left\{   \bI^{-1}(\xi_{\phi}^{*}, \btheta^*)\right\}$ as 
$n \to \infty$,
which gives
$\lim_{n \to \infty} \Eff_{\phi}(\xi_{n,\phi}^*) =1$.

\bigskip

For part (ii),  we want to show that 
$\lim_{N \to \infty} \Eff_{\phi}(\xi_{\phi, N}^*) =1$.

For each $S_N$, we construct an approximate design $\xi_N^{(0)}$
with $m$ support points, which are selected as follows. 
By the construction of $S_N$, we can find a sequence of points,
${\bf u}_{i,N} \in S_N$, $i=1, \ldots, m$, such that  
$\lim_{N \to \infty}  {\bf u}_{i,N} =\bx_i^*$, for all $i=1, \ldots, m$.
Then we choose ${\bf u}_{i,N} \in S_N$, $i=1, \ldots, m$, as the support points of $\xi_N^{(0)}$,
and their weights are 
 $w_1^*, \ldots, w_m^*$ (from $\xi_{\phi}^*$), respectively.
The information matrix of the approximate design $\xi_N^{(0)}$ using (\ref{InfoG}) is, 
\begin{eqnarray*}
\bI(\xi_N^{(0)}, \btheta^*) &=& \sum_{i=1}^m  w_i^*
~I({\bf u}_{i,N}, \btheta^*) \\
& \to & \sum_{i=1}^m  w_i^*
~I(\bx_{i}^*, \btheta^*), ~\mbox{as}~N \to \infty \\ 
&=& \bI(\xi_{\phi}^{*}, \btheta^*),
\end{eqnarray*} 
since  all entries of $I(\bx, \btheta^*)$ are continuous functions of
$\bx$.  Similar to the proof of part (i),
it is clear that 
$\phi \left\{   \bI^{-1}(\xi_N^{(0)}, \btheta^*) \right\}  \to
\phi \left\{   \bI^{-1}(\xi_{\phi}^{*}, \btheta^*)\right\}$ as 
$N \to \infty$.
Since $\xi_N^{(0)}$ is a design on $S_N$ and $\xi_{\phi,N}^*$ is an OAD
on $S_N$, it follows that
$$\phi \left\{   \bI^{-1}(\xi_N^{(0)}, \btheta^*) \right\}  \ge
\phi \left\{   \bI^{-1}(\xi_{\phi,N}^*, \btheta^*) \right\} 
\ge \phi \left\{   \bI^{-1}(\xi_{\phi}^{*}, \btheta^*)\right\}.$$
Thus, 
$\phi \left\{   \bI^{-1}(\xi_{\phi,N}^{*}, \btheta^*) \right\}  \to
\phi \left\{   \bI^{-1}(\xi_{\phi}^{*}, \btheta^*)\right\}$ as 
$N \to \infty$,
which implies that 
$\lim_{N \to \infty} \Eff_{\phi}(\xi_{\phi, N}^*) =1$.
\hfill{$\Box$}

\bigskip

\section*{Acknowledgements}
This research was partially supported by Discovery Grants from the Natural Sciences and Engineering Research Council of Canada. The first author is partially supported by the CANSSI Distinguished Postdoctoral Fellowship from the Canadian Statistical Sciences Institute.

\section*{Declarations}
The authors have no conflicts of interest to declare.

\linespread{1}\small
\bibliographystyle{apalike} 
 \bibliography{CVXSADes_arxiv}
\end{document}